\documentclass[useAMS,usenatbib]{mn2e}
\usepackage{times}
\usepackage{graphicx}

\newcommand{\kms}   {km~s$^{-1}$}
\newcommand{\hto}   {H$_2$O}		
\newcommand{\jpb}   {$\mbox{Jy~beam}^{-1}$}

\newcommand{\eg}    {e.\,g.}
\newcommand{\ie}    {i.\,e.}
\newcommand{\hii}   {H\,{\sc ii}}
\newcommand{\sii}   {[S\,{\sc ii}]}
\newcommand{\nii}   {[N\,{\sc ii}]}
\newcommand{\fe}    {[Fe\,{\sc ii}]}
\newcommand{\vlsr}  {$V_\rmn{LSR}$}
\newcommand{\iras}  {\textit{IRAS}}
\newcommand{\tmas}  {\textit{2MASS}}
\newcommand{\nir}   {near-\textit{IR}}
\newcommand{\fir}   {far-\textit{IR}}

\title[Integral Field Spectroscopy of HHL~73]{
The nature of HHL~73 from optical  imaging and Integral Field Spectroscopy
}

\author[R. L\'opez et al.]
{R. L\'opez,$^{1}$\thanks{E-mail:
\mbox{rosario@am.ub.es};
\mbox{sanchez@caha.es};
\mbox{bgarcia@iac.es};
\mbox{gabriel.gomez@gtc.iac.es};
\mbox{robert.estalella@am.ub.es};
\mbox{angels.riera@upc.edu};
\mbox{gbusquet@am.ub.es}}
S. F. S\'anchez,$^{2}$\footnotemark[1] 
B. Garc\'\i a-Lorenzo,$^{3}$\footnotemark[1]
G. G\'omez,$^{3,4}$\footnotemark[1]
\newauthor
R. Estalella,$^{1}$\footnotemark[1] 
A. Riera$^{5}$\footnotemark[1] and 
G. Busquet$^{1}$\footnotemark[1]
\thanks{Based on observations collected at the Centro Astron\'omico Hispano
Alem\'an (CAHA) at Calar Alto, operated jointly by the Max-Planck Institut
f\"ur Astronomie and the Instituto de Astrof\'{\i}sica de Andaluc\'{\i}a (CSIC)
and in the 2.6 m Nordic Optical Telescope and 2.5 m Isaac Newton Telescope at
the Observatorio del Roque de los Muchachos of the Instituto de
Astrof\'{\i}sica de Canarias.
}\\
$^{1}$Departament d'Astronomia i Meteorologia, Universitat de Barcelona,
Mart\'{\i} i Franqu\`es 1, E-08028 Barcelona, Spain.\\
$^{2}$Centro Astron\'omico Hispano-Alem\'an de Calar Alto, Jes\'us Durban Remon
2-2, E-04004 Almer\'{\i}a, Spain.\\
$^{3}$Instituto de Astrof\'{\i}sica de Canarias, E-38200 La Laguna, Spain\\
$^{4}$GTC Project Office, GRANTECAN S.A. (CALP), E-38712 Bre\~na Baja, La
Palma, Spain.\\
$^{5}$Dept.\ F\'{\i}sica i Enginyeria Nuclear. EUETI de Barcelona. Universitat
Polit\`ecnica de Catalunya. Compte d'Urgell 187, E-08036 Barcelona, Spain.
}

\begin{document}

\date{Accepted. Received;}

\pagerange{\pageref{firstpage}--\pageref{lastpage}} \pubyear{2007}

\maketitle

\label{firstpage}

\begin{abstract}
We present new results on the nature of the Herbig--Haro-like object 73 
(HHL~73, also known as [G84b]~11) based on narrow-band CCD H$\alpha$ and \sii\
images of the HHL~73 field, and Integral Field Spectroscopy and radio
continuum  observations at 3.6 cm covering the emission of the  HHL~73 object.
The CCD images allow us to resolve the HHL~73 comet-shaped  morphology
into two components and a collimated emission feature of $\sim4$-arcsec long,
reminiscent of a microjet. The IFS spectra of HHL~73 showed emission
lines characteristic of the spectra of Herbig--Haro objects. The kinematics
derived for HHL~73 are complex. The profiles of the \sii\ $\lambda\lambda$6717,
6731 \AA\ lines were well fitted with a model of three Gaussian velocity
components peaking at $\mbox{\vlsr}\simeq-100$, $-20$ and +35 \kms. We found
differences among the spatial distribution of the kinematic components that are
compatible with the emission from a bipolar outflow with two blueshifted 
(low- and high-velocity) components. Extended radio continuum emission at
3.6 cm was detected showing a distribution in close agreement with the HHL~73
redshifted gas.
From the results discussed here, we propose HHL~73 to be a true HH object.
\iras~21432+4719, offset 30 arcsec northeast from the HHL~73 apex, is the most
plausible candidate to be driving HHL~73, although the evidence is not
conclusive.
\end{abstract}


\section{Introduction}

Herbig--Haro-like objects (HHLs) appear as bright nebulosities in the red
optical images. Spectroscopy of HHLs indicates that a fraction of them are
reflection nebulae originated by the scattered light from a stellar source
associated with the nebulosity, while only a fraction of HHLs are Herbig--Haro
(HH) objects arising from shock-excited gas. 
HHL~73, also known as [G84b]~11, was first reported by \citet{Gyu84} as 
object 11 found in his survey of \textit{DSS} red plates. In these plates, the
object shows a comet-shaped morphology, being located close to the western
border of a region with high visual extinction. In fact, HHL~73 is found
associated with a  dark cloud complex in Cygnus, close to the Lynds Dark Cloud
1035A (\textit{LDN} 1035A), and lying in a region with clear signs of recent
star formation activity. The estimated distance to the region is
$900\pm100$~pc \citep{Eli78}.

The source \iras~21432+4719 is located  close to HHL~73. Both sources are
engulfed within a clump of high-density molecular gas, traced by the emission
of the NH$_3$ and CS molecules, whose emission peaks around the HHL~73 and
\iras~21432+4719 positions (see \citealt{Ang97} and references therein for a
wider description on the high-density molecular gas emissions in the HHL~73
field). 
Emission from high-velocity molecular gas in this field is also reported  by
\citet{Dob93}, who detect a highly asymmetric CO bipolar outflow. The emission
of the redshifted outflow lobe is spatially more extended than the blueshifted
lobe and encompasses the region around \iras\ and HHL~73. In addition, the
detection of a water maser near HHL~73 was reported by \citet{Gyu87}. 

\begin{figure}
\centering
\includegraphics[width=\hsize]{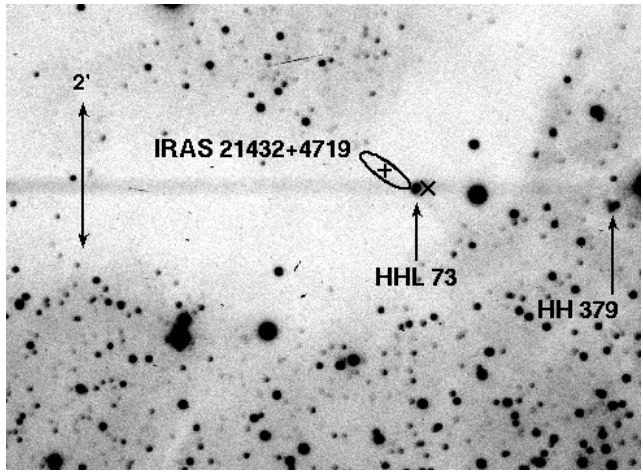}
\caption{
CCD H$\alpha$-band image of the HHL~73 field obtained at the Isaac Newton 
Telescope (INT, ORM). Arrows point to the HHL~73 and HH~379 objects. The
position of \iras~21432+4719 and the error ellipse are marked. The tilted
cross marks the position given by \citet{Gyu87} for the H$_2$O maser. The image
FOV is $\sim8.5\times6$ arcmin$^2$. North is up and east is to the left.
\label{fig1}}
\end{figure}

\citet{Dev97} imaged the HHL~73 field through narrow-band H$\alpha$ and \sii\
$\lambda\lambda$6717, 6731 \AA\ filters. They find two stellar-like
condensations $\sim$ 2.5 arcmin west of HHL~73 with similar H$\alpha$ and \sii\
surface brightness, surrounded by diffuse emission. They designate their
finding as the new Herbig--Haro object HH 379. In addition, they suggest that
the exciting source of HH~379 may be embedded within the cometary nebula which
appears in their images at a position  close to \iras~21432+4719 (thus, at
the position given by \citet{Gyu84} for HHL~73). These authors also discuss a
plausible association between the \iras\ source  and the undetected  exciting
source of HH~379. It should be noted that  \citet{Dev97} never refer to the
optical cometary nebula as the HHL~73 object reported by \citet{Gyu84}. Later
on, \citet{Dav01, Dav03} performed \nir\ echelle spectroscopy at the HHL~73
position. They detect emission  from H$_2$ $\lambda$2.122 $\mu$m and \fe\
$\lambda$1.644 $\mu$m lines, while no continuum  emission is detected at the
HHL~73 position. Both lines show a rather complex profile, the width of the
\fe\ line being roughly twice that of the H$_2$ line. \citet{Dav03} suggest that
the \fe\ emission may trace the higher-velocity component of a jet. In
addition, \citet{Dav01, Dav03} also suggest that the optical nebula  found in
the \citet{Dev97} images harbours the true exciting source of HH~379, thus
discarding the possibiility that \iras~21432+4719 was driving HH~379. This yet undetected
exciting source of HH~379 is named as HH 379-IRS by these authors. In order to
help with the identification of the sources mentioned here, we show in
Fig.\ \ref{fig1} a CCD H$\alpha$ image of the HHL~73 field where all these
sources have been marked.

With the aim of unravelling the nature of HHL~73, we included this target
within a programme of Integral Field Spectroscopy (IFS) of Herbig--Haro objects
using the  Potsdam Multi-Aperture Spectrophotometer (PMAS) in the wide-field
IFU mode PPAK. Additional CCD narrow-band images and radio continuum 3.6 cm
data have been analysed to complement the study. We present in this paper the
results on the kinematics and physical conditions of HHL~73 obtained from this
IFS pilot  programme.

The paper is organized as follows. The observations and data reduction of CCD
imaging, radio continuum and IFS observations  are described in \S\ 2.1, 2.2
and 2.3 respectively. Results are given in \S\ 3: CCD imaging in \S\ 3.1, radio
continuum in \S\ 3.2 and IFS in \S\ 3.3, including results from integrated
emission (\S\ 3.4) and a line-profile decomposition model for \sii\  profiles
(\S\ 3.5). A global discussion is made in \S\ 4. In \S\ 5 the main conclusions
are summarized.

\section{Observations and data reduction}

\subsection{Narrow-band CCD images}

\begin{figure*}
\centering
\includegraphics[width=0.4\hsize,clip]{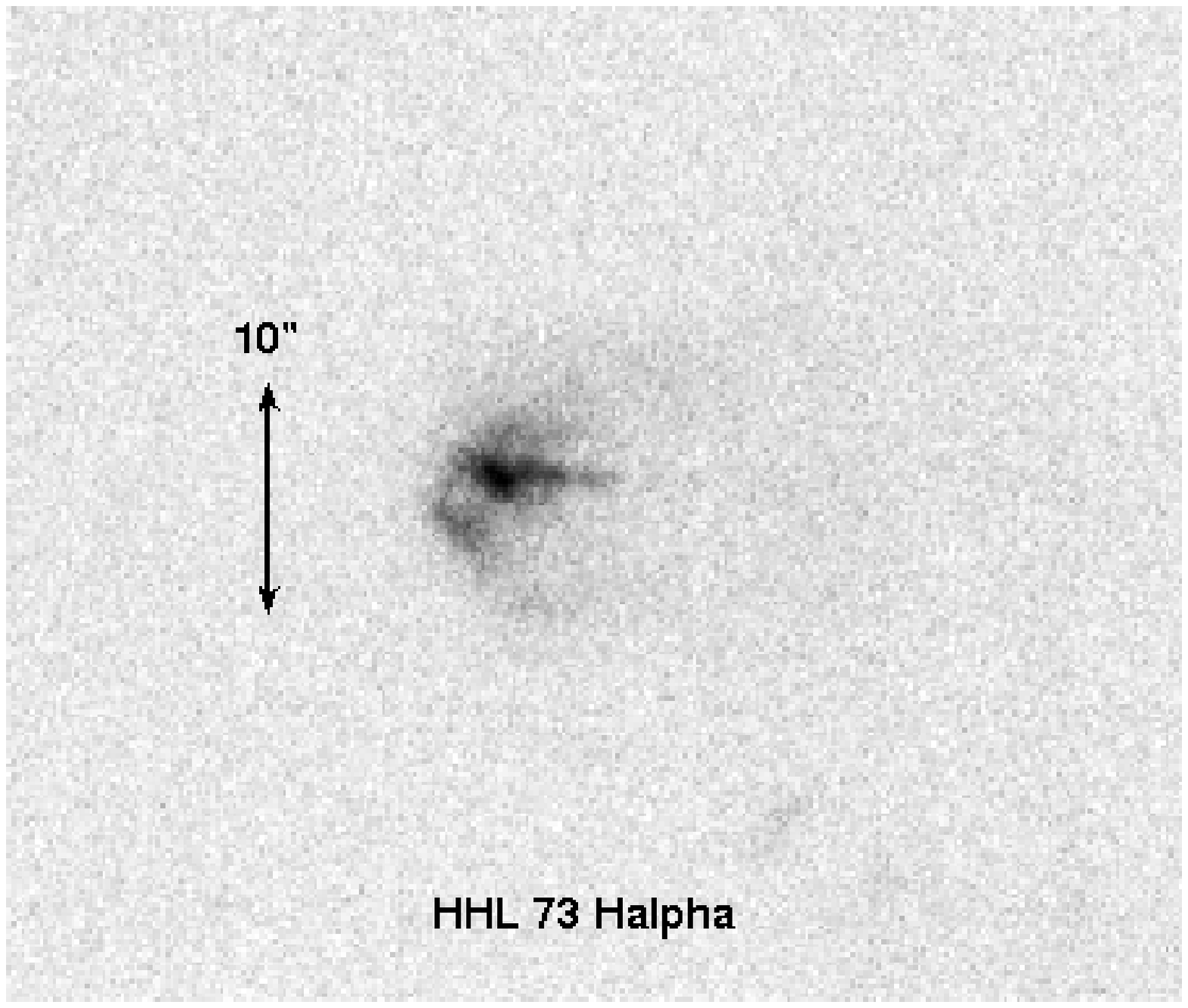}%
\includegraphics[width=0.4\hsize,clip]{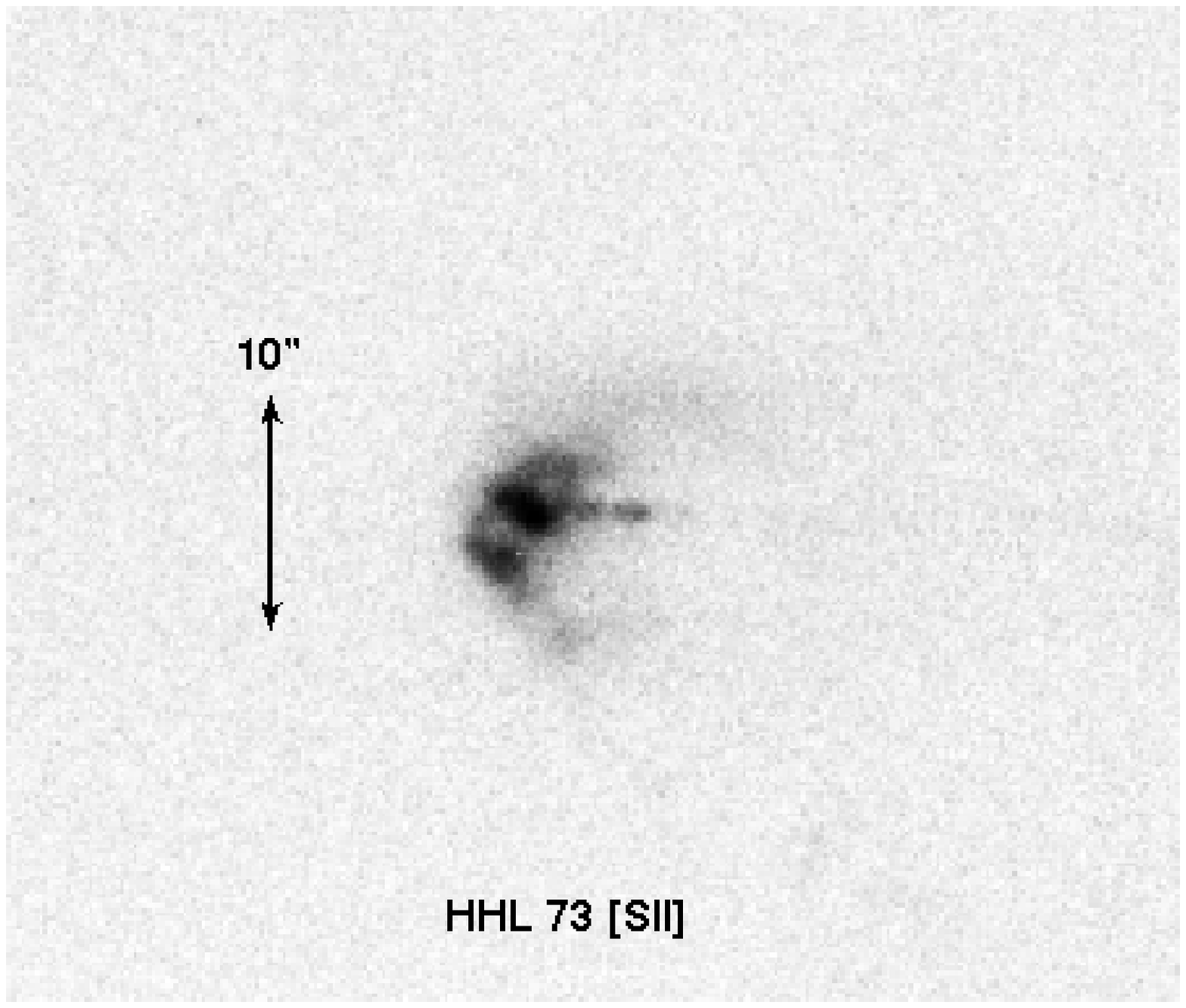}
\caption{
Close-up of the HHL~73 images obtained at the Nordic Optical Telescope (NOT,
ORM) through the line filters of  H$\alpha$ (\emph{left}) and   \sii\
(\emph{right}). In both images, the spatial structure of HHL~73 emission is
resolved into two components. A microjet of $\sim4$ arcsec is also visible in
the  two band images. The FOV of each image is $\sim50\times42$ arcsec$^2$.
North is up and east is to the left.
\label{fig2}}
\end{figure*}

Narrow-band CCD images of the HHL~73 field were made on 2006 August 5 at the
2.6 m Nordic Optical Telescope (NOT) (ORM, La Palma, Spain) using the Service
Time facility. The images were obtained  with the Andalucia  Faint Object
Spectrograph and Camera (ALFOSC). The  image scale was 0.188
arcsec~pixel$^{-1}$. The  effective imaged field was $\sim~5\times5$ arcmin$^2$.
Images were obtained through filters of H$\alpha$ (central wavelength,
$\lambda_c=6564$~\AA, bandpass $\Delta\lambda=33$~\AA, which mainly includes
emission from this line, although some contamination from \nii\ emission is also
present) and \sii\   ($\lambda_c=6725$~\AA, $\Delta\lambda=60$~\AA), which
includes the two red \sii\ lines. The  integration time was 1800 s in H$\alpha$
and 2400 s in \sii. The seeing was 0.6--0.8 arcsec. 
In order to check whether there is significant continuum contribution to the
HHL~73 red emission, additional exposures (600 s integration time) were acquired through a narrow-band,
off-line filter, but no continuum signal around the HHL~73
location was detected.
All the images
were processed with the standard tasks of the {\small IRAF}%
\footnote{{\small IRAF} is distributed by the National Optical Astronomy
Observatories, which are operated by the Association of Universities for
Research in Astronomy, Inc., under cooperative agreement with the National
Science Foundation.} 
reduction package. Astrometric calibration of the two final images  was
performed.  In order to register the images, we used the  coordinates  from the
\textit {USNO} Catalogue%
\footnote{The {\textit {USNOFS}} Image and  Catalogue Archive is operated by the
United States Naval Observatory, Flagstaff Station.} 
of eight field stars well distributed on the  observed field. The rms of the
transformation was 0.12 arcsec in both coordinates.\\

In addition to the NOT images, we also show in Fig.\ \ref{fig1} a CCD image of
the HHL~73 field obtained on 2003 December 15 at the 2.5 m Isaac Newton
Telescope (INT) through a narrow-band filter of central wavelength
$\lambda_c=6564$~\AA\ and bandpass $\Delta\lambda=50$~\AA\ that  also includes
emission from the \nii\ $\lambda\lambda$6548, 6583~\AA\ lines. The  integration
time was 3600 s. Astrometric calibration of this image was also performed with
an accuracy of a fraction of the pixel size (0.24 arcsec rms in both
coordinates). This image has a spatial resolution significantly lower than that
of the NOT images since the image scale was 0.54 arcsec~pixel$^{-1}$ and, in
addition, it was obtained under worse seeing conditions ($\ge3$ arcsec).
However, the INT image is deeper than the NOT images and has been used in our
study to look for faint emission nebulae in the field that might be related to
the HHL~73 object and/or with HH~379.

\subsection{Radio continuum and H$_2$O maser emission}

We observed the 3.6 cm continuum emission towards HHL~73 with the Very Large
Array of the NRAO\footnote
{The National Radio Astronomy Observatory is operated by Associated Universities
Inc., under cooperative agreement with the National Science Foundation.} 
in the C configuration during 1997 August 14. The phase centre was located at
$\rmn{RA}(J2000)=21^{\rm h}45^{\rm m}10\fs46$; 
$\rmn{Dec.}(J2000)=+47\degr33\arcmin21 \farcs2$. 
The phase calibrator used was 2202+422, and the bootstrapped flux density
obtained was $3.69\pm0.07$~Jy. The absolute flux density calibrator was
00137+331 (3C48), for which a flux density of 3.15~Jy was adopted. Clean maps
were obtained using the task {\small MAPPR} of {\small AIPS} with the robust
parameter set equal to 5, which is close to natural weighting. Since the
signal-to-noise ratio of the longest baselines was low, we applied a
$(u,v)$-taper function of 50~k$\lambda$. The resulting synthesized beam size
was  $4.7\times4.3$ arcsec$^2$ ($\rmn{P.A.}=-78\degr$), and the achieved
$1\sigma$ noise level was 20~$\mu$\jpb.
 
VLA archive observations of the water maser line at 22.2351~GHz 
($6_{16}$--$5_{23}$ transition) carried out in the B configuration during 1994
July 4 were analysed. The phase centre was located at 
$\rmn{RA}(J2000)=21^{\rm h}45^{\rm m}08\fs37$;
$\rmn{Dec.}(J2000)=+47\degr32\arcmin49\farcs1$. 
The bootstrapped flux obtained for the phase calibrator 2202+422 was
$3.08\pm0.09$~Jy, and the adopted flux density for the absolute flux calibrator
(3C48) was 1.13~Jy at 1.3~cm. The observations were made using the 2IF mode,
with a bandwidth of 6.30~MHz with 31 channels of 203~kHz resolution (2.66~\kms\
at 1.3~cm) centred at $-2$~\kms, plus a continuum channel that contains the
central 75 percent of the bandwidth. The synthesized beam was $0.36\times0.31$
arcsec$^2$ ($\rmn{P.A.}=-11\degr$), and the $1\sigma$ noise level per
spectral channel achieved with natural weight was 8~m\jpb.

\begin{figure*}
\centering
\includegraphics[width=0.4\hsize]{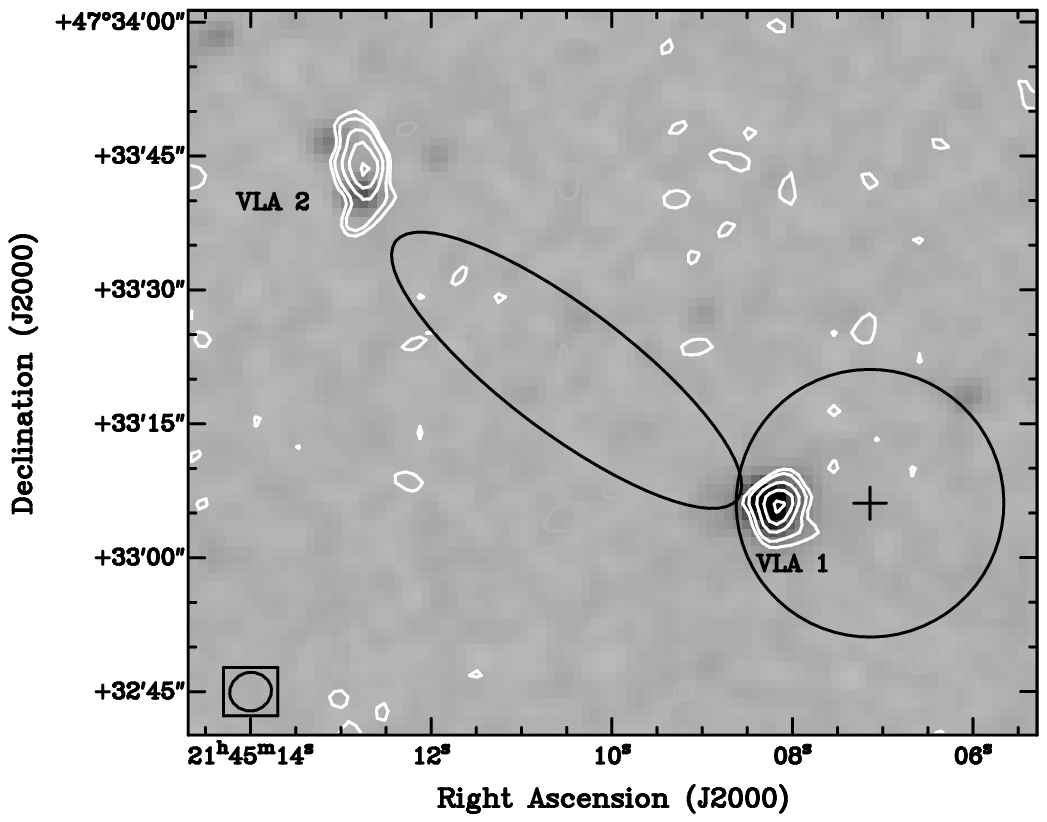}%
\includegraphics[width=0.4\hsize]{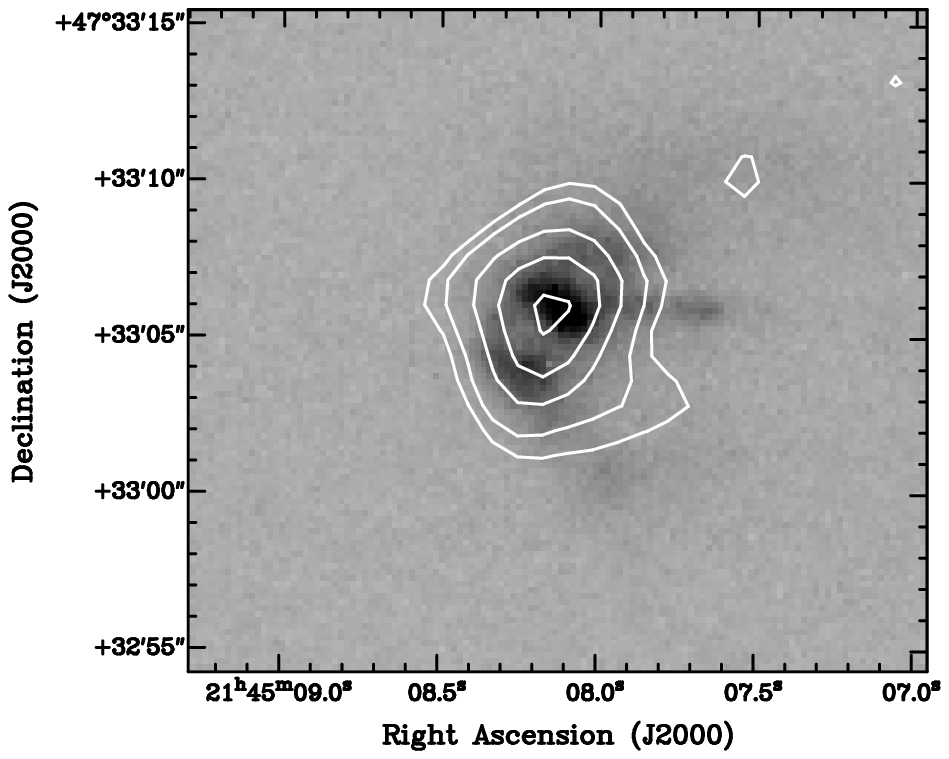}
\caption{
\emph{Left panel}: Continuum emission at 3.6~cm (white contours)  overlaid on
the \textit{K}-band \tmas\ image (grey scale). The cross marks the  position of the
\hto\ maser detected by \citet{Gyu87}. The synthesized beam is shown at the
bottom left corner, and the black  ellipse indicates the \iras\ error ellipse.
\emph{Right panel}: Detail of the 3.6~cm continuum emission towards HHL~73
(white contours) overlaid on the \sii\ image (grey scale). In both panels,
contours levels are $-5$, $-3$, 2, 3, 5, 7, and 9 times the rms of the map,
20~$\mu$\jpb.
\label{fig3}}
\end{figure*}

\subsection{Integral Field Spectroscopy (IFS)}

Observations of the HHL~73 object were made on 2004 November 22 with the 3.5 m
telescope of the Calar Alto Observatory (CAHA). Data were acquired with the 
Integral Field Instrument Potsdam Multi-Aperture Spectrophotometer PMAS
\citep{Rot05} using the PPAK configuration that has 331 science fibres, covering
a hexagonal FOV of 74$\times$65 arcsec$^2$ with a spatial sampling of 2.7
arcsec per fibre. In addition, PPAK comprises 36 fibres to sample the sky,
located $\sim90$ arcsec from the centre, packed in small bundles of six fibres
each one, and distributed at the edges of the science hexagon. Fifteen
calibration fibres are interpolated in between the science and sky ones in the
pseudo-slit, and connected directly to the PMAS internal calibration unit
\citep{Ke06}. The I1200 grating was used, giving an effective sampling  of 0.3
\AA\ pix$^{-1}$  ($\sim15$  km~s$^{-1}$ for H$\alpha$) and covering the
wavelength range $\sim6500$--7000 \AA, thus including characteristic HH emission
lines in this wavelength range (H$\alpha$, \nii\ $\lambda\lambda$6548, 6584
\AA\  and \sii\  $\lambda\lambda$6717, 6731 \AA). The spectral resolution (\ie\
instrumental profile) is $\sim2$ \AA\  FWHM ($\sim90$ km~s$^{-1}$) and the
accuracy in the determination of the position of the line centroid is $\sim0.2$
\AA\  ($\sim10$ km~s$^{-1}$ for the strong observed emission lines). We made a
single pointing of 1800-s exposure time centred on the \textit{SIMBAD}
position of HHL~73  
($\rmn{RA}(J2000)=21^{\rmn{h}}45^{\rmn{m}}08\fs1$, 
$\rmn{Dec.}(J2000)=+47\degr33\arcmin6\arcsec$)    
covering the whole optical emission of the object ($\sim30$ arcsec).

Data reduction was performed using a preliminary version of the {\small R3D}
software \citep{Sa06}, in combination with {\small IRAF}  and the {\small
Euro3D} packages \citep{Sa04}. The reduction consists of the standard steps for
fibre-based integral field spectroscopy. A master bias frame was created by
averaging all the bias frames observed during the night and subtracted from the
science frames. The location of the spectra in the CCD was determined using a
continuum-illuminated exposure taken before the science exposures. Each spectrum
was extracted from the science frames by co-adding the flux  within an aperture
of five pixels along the cross-dispersion axis for each pixel in the dispersion
axis, and stored it in a row-stacked-spectrum (RSS) file \citep{Sa04}. At the
date of the observations we lacked a proper calibration unit for PPAK.
Therefore, the wavelength calibration was performed  iteratively. First,
we illuminated the science fibres using dome  \textit{strahler} lamps with the
telescope at the location of the objects. These lamps have well defined emission
lines in the wavelength range considered, although the line elements and
wavelength are ignored at this step. These exposures  were used to correct for
distortions in the wavelength solution, fixing all the spectra to a common
(ignored) solution. After that, the wavelength solution was found using HgNe
lamps that illuminated only the calibration fibres. Differences in the
fibre-to-fibre transmission throughput were corrected by comparing the
wavelength-calibrated RSS science frames with the corresponding continuum
illuminated ones. The final wavelength solution was set by using the sky
emission lines found in the observed wavelength range. The achieved accuracy 
for the wavelength calibration was better than  $\sim0.1$  \AA\ ($\sim5$
km~s$^{-1}$).
Observations of a standard star were used to perform a relative flux calibration.
A final datacube  containing the 2D spatial plus the spectral information of
HHL~73 was then created from the 3D data by using {\small Euro3D} tasks  to
interpolate the data spatially  until reaching a final grid of 2.16 arcsec of
spatial resolution and 0.3 \AA\ of spectral resolution. Further manipulation of
this datacube, devoted to obtaining  integrated emission line maps, channel maps
and position--velocity maps, were made using several common-users tasks of {\small
STARLINK}, {\small IRAF} and {\small GILDAS} astronomical packages and {\small
IDA} \citep{Ga02}, a specific suit of {\small IDL} software for analysing 3D data.

\section{Results} 

\subsection{Narrow-band CCD images}

Figure \ref{fig1} shows the H$\alpha$ image of the HHL~73 field obtained at the
INT. The HHL~73 object appears projected on a dark cloud near its southwestern
border. The object presents in this  H$\alpha$ image the comet-shaped
morphology already found in the \textit{DSS} plates. Because of the low
effective spatial resolution of the image resulting from both the pixel scale
and the seeing, we were unable to resolve any structure in the morphology of
the H$\alpha$ emission. Neither were we able to elucidate from this
image whether the emission of HHL~73 originates mainly from scattered
starlight from an embedded point-source, as suggested by \citet{Dev97}, or
whether  the emission mainly arises from gas photoionized or shock-excited by
an external source. 
The error ellipse of \iras~21432+4719 does not encompass HHL~73. From the
astrometry performed on the H$\alpha$ image, we derived that the  \iras\
catalogue position is $\sim30$  arcsec northeast of the HHL~73 H$\alpha$
emission peak ($\sim0.13$ pc for a distance of 900 pc).  
A \tmas\ source is found very close to HHL~73 (there is an offset of
$\sim1.5$ arcsec between the position given for the infrared source in the
\tmas\ catalogue and the position of the H$\alpha$ emission peak of HHL~73).
The nebulosities cited by \citet{Dev97} as HH~379 are also visible in our INT
image close to a bright field star (see Fig.\ \ref{fig1}). However, despite a
careful inspection of the HHL~73 field imaged in H$\alpha$, we were unable to
identify any other nebular emission suggestive of being part of a putative
optical jet related to HH~379.

\begin{table*}
\begin{minipage}{110mm}
\caption{Parameters of the continuum sources detected in the HHL~73 region.
\label {tjcmt}}
\begin{tabular}{@{}lccccc@{}}
\hline
& \multicolumn{2}{c}{Position} 
& Flux density\footnote{Corrected for primary beam response.}
& Deconv.\ size & P.A.\\
\cline{2-3}
Source & $\alpha (J2000)$ & $\delta (J2000)$ & (mJy) & (arcsec)& (deg)\\
\hline
VLA 1 (HHL~73)&  21$^{\rm h}$45$^{\rm m}$08$\fs15$& 
47$\degr$33$\arcmin$05$\farcs$6& 0.26$\pm$0.02&	 3.6$\times$2.0&   170\\
VLA 2&	 21$^{\rm h}$45$^{\rm m}$12$\fs77$& 
47$\degr$33$\arcmin$43$\farcs$6& 0.33$\pm$0.03&	 7.5$\times$0.0&	   9\\ 
\hline
\end{tabular}
\end{minipage}
\end{table*}

The narrow-band H$\alpha$ and \sii\ images obtained at the NOT have less depth
than the INT image. However the higher spatial resolution achieved in the NOT
images make them suitable for resolving the structure of the HHL~73 object. 
Fig.\ \ref{fig2} shows the close-up of the NOT images including the HHL~73
object in the H$\alpha$ (right) and \sii\ (left) bands. As can be seen from the
figure, these images allow us to resolve several components in the emission of HHL~73:
the comet-shaped morphology detected in the optical INT images of 
lower spatial resolution now appears split into two components of different
surface brightness with a darker lane between them. The estimated distance
between the peaks of the two HHL~73 components is $\sim2.5$ arcsec ($\sim0.01$
~pc for a distance of 900~pc). In addition, a lower brightness and more
collimated emission tail, showing a couple of brightness enhancements
suggestive of the ``knots'' of a microjet, emanates from the brighter
northwestern component and extends $\sim4$ arcsec towards the west at
$\mbox{P.A.}=267\degr$. All these HHL~73 components are found in both NOT
images. However, some differences are appreciated when the emission from
the two bands is compared: the northwestern component is the brighter one in
both filters, H$\alpha$ and \sii. The southeastern component appears dimmer, as
compared with the northwestern one, in H$\alpha$ than in \sii. The emission
tail is brighter and has higher collimation in \sii\ than in H$\alpha$ (as
would be expected from a jet). Furthermore, while the emission peak positions
of H$\alpha$ and \sii\ are coincident in the southeastern component, the
H$\alpha$ peak position in the northwestern component is offset by $\sim0.8$
arcsec towards the southwest from the \sii\ emission peak position. Thus,
differences in the excitation conditions between the two HHL~73 components
might be present.

Finally, note that the position angle of HH 379, as seen from HHL~73, is 
$262\degr$ \citep{Dav01, Dav03}, close to that of the HHL~73 microjet. Thus,
although the deep INT H$\alpha$ image does not show any other nebulosity
connecting both emissions, HH 379 could form part of the HHL~73 optical
outflow, provided that the outflow direction undergoes a bending of
$\sim5\degr$, as seen in other optical outflows propagating in a inhomogeneous
medium \citep{Lop95}.

\subsection{Centimetre continuum and \hto\ maser emission}

In order to complement the CCD and IFS data of HHL~73, we analysed the 
data of the emission at radio wavelengths available for the HHL~73 field.

\subsubsection{Centimetre continuum emission}

Figure~\ref{fig3} (left panel) shows the continuum map at 3.6 cm towards HHL~73 
overlaid on the \textit{K}-band \tmas\ image. We detected two centimetre sources,
VLA~1 and VLA~2, separated $\sim47$ arcsec (0.2~pc in projection at the
distance of the source), located near the catalogue position of
\iras~21432+4719. Both centimetre sources are also associated with \tmas\ point
sources. In Table 1 we list the positions, flux densities and deconvolved sizes
of the detected sources in the HHL~73 region. In Fig.\ \ref{fig3} (right panel)
we show the 3.6 cm centimetre continuum emission arising from the HHL~73 object
overlaid on the \sii\  image. The peak position of the centimetre source VLA~1
is close ($\sim0.7$ arcsec south) to  the position of the \sii\ emission peak.
The centimetre emission from the HHL~73 object is slightly elongated towards
the secondary peak detected in the \sii-band images to the southeast.

\subsubsection{\hto\ maser emission}

In 1985 \hto\ maser emission was detected towards HHL~73 by \citet{Gyu87} using
the 37 m radio telescope of the Haystack Observatory. The \hto\ maser was
located 10.2 arcsec west of the HHL~73 object, with a positional uncertainty
that can be estimated at $\sim$15 arcsec, making it difficult to ascertain
whether or not the maser is associated with HHL~73. In the archive observations
analysed here we did not detect \hto\ maser emission towards HHL~73 above a
$6\sigma$ limit, indicating that at the epoch of observation (1994 July) the
maser was out. The upper limit estimated for the peak intensity of the \hto\
maser emission was 48~m\jpb, corresponding to a main beam brightness
temperature upper limit $T_\mathrm{MB}=1000$~K.

\subsection{IFS: Integrated line-emission maps}

\subsubsection{Emission line images}

\begin{figure}
\centering
\includegraphics[angle=-90,width=\hsize]{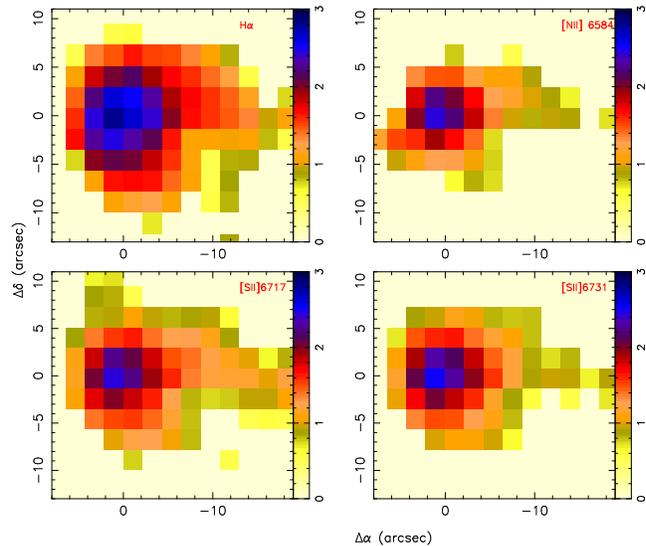}
\caption{
Integrated maps of the HHL~73 emission obtained from the datacube  by
integrating the signal within the wavelength range including the corresponding
line labelled in each panel 
(H$\alpha$: $\lambda$6557.7--6569.1 \AA, 
\nii:       $\lambda$6579.6--6585.9 \AA, 
\sii $\lambda$6716:  6712.1--6719.3 \AA\  and 
\sii $\lambda$6731:  6728.0--6733.8 \AA).  
Fluxes  are displayed on a logarithmic scale, in  units of $10^{-16}$
erg~s$^{-1}$~cm$^{-2}$. The offsets of the spatial scale are relative to the
position of the emission peak. 
\label{fig4}}
\end{figure}

We detected emission from characteristic HH lines within the  observed
wavelength range (H$\alpha$, \nii\ $\lambda\lambda$6548, 6584 \AA\  and \sii\
$\lambda$6717, 6731 \AA). Narrow-band images for these lines were generated
from the datacube. For each position, the flux of the emission line was
obtained by integrating the signal over the  whole wavelength  range covered by
the line and subtracting a continuum obtained from the wavelengths adjacent to
the emission line. 

The narrow-band maps of HHL~73 obtained are shown in Fig.\ \ref{fig4}. As  can
be seen from the panels of the figure, the emission has a similar morphology in
all the mapped lines. However it should be noted that the H$\alpha$ emission is
slightly more extended ($\sim3$ arcsec) towards the south than the emission
from the other lines.  Furthermore, the \nii\ emission appears more compact
than the H$\alpha$ and \sii\ emission.  This is mostly because the  \nii\
emission has a lower surface brightness, and the collimated emission extending
towards the west of the HHL~73 main body is hardly detected in the \nii\
lines. 

It should also be noted that the overall morphology derived from the H$\alpha$
and \sii\ IFS images is in good agreement with the morphology found from the
INT CCD image. In particular, the IFS images also show the low-brightness and
more collimated emission feature originating west of the HHL~73 main body.
However, the lower spatial resolution of the IFS images did not allow us to
resolve the two components of the HHL~73 main body discovered from the 
H$\alpha$ and \sii\ NOT images.

\subsubsection{Kinematics}

\begin{figure}
\centering
\includegraphics[angle=-90,width=\hsize]{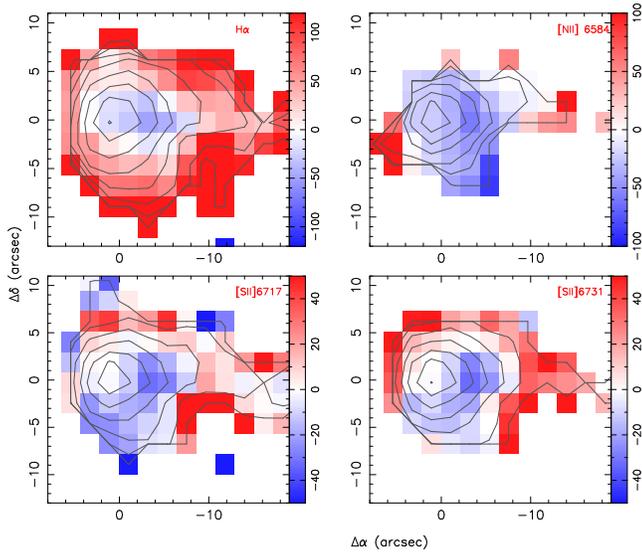}
\caption{
Flux-weighted mean velocity maps. Velocities (km~s$^{-1}$) are referred to the
\vlsr\ of the parent cloud. The offsets of the spatial scale are relative to
the position of the emission peak. Contours of the integrated flux of the
corresponding line have been overlaid. The lowest contour level is 
$7\times10^{-17}$ erg~s$^{-1}$~cm$^{-2}$  and the contour spacing is
$3\times10^{-17}$ erg~s$^{-1}$~cm$^{-2}$.
\label{fig5}}
\end{figure}

In addition to the integrated flux, we derived for each position the
flux-weighted mean radial velocity\footnote{
All the velocities in the paper are referred to the local standard of rest
(LSR) frame. A   $\mbox{\vlsr}=+3.6$ \kms\ for the parent cloud has been taken
from \citet{Dob93}.} 
$\langle V\rangle$, and the flux-weighted rms width of the line $(\Delta
V^2)^{1/2}$, proportional to the FWHM, which are given by
\begin{equation}
\langle V\rangle=\frac{1}{I}\int\limits_\rmn{line} I_V V\,\rmn{d}V
\end{equation}
and
\begin{equation}
\Delta V^2=
\frac{1}{I}\int\limits_\rmn{line} I_V (V-\langle V\rangle)^2\,\rmn{d}V,
\end{equation}
where 
\begin{equation}
I=\int\limits_\rmn{line} I_V\,\rmn{d}V
\end{equation}
In spite of the spatial extension of the emission being quite compact, the
velocity field derived for $\langle V\rangle$ appears rather complex for all 
the lines and shows appreciable changes through HHL~73, as  can be seen from
Figure \ref{fig5}.

For all the mapped lines, the gas appears blueshifted around the emission peak,
with $\langle V\rangle$ values of $-20$ km~s$^{-1}$ for H$\alpha$ and
\nii, and $-2.0$ km~s$^{-1}$ for the \sii\ lines. However, the highest
blueshifted velocities are found $\sim5$ arcsec west of the emission peak
position, where values of $-50$~km~s$^{-1}$ for H$\alpha$ and \nii, and 
$-32.0$ km~s$^{-1}$ for the \sii\ lines have been derived.

The FWHM values obtained from the $(\Delta V^2)^{1/2}$, corrected from the
instrumental profile, are high for all the mapped lines. Around the emission
peak position, FWHMs are $\sim190$ km~s$^{-1}$ for the H$\alpha$ line and
$\sim100$ km~s$^{-1}$ for the \nii\ and \sii\ lines. These values are
representative of the FWHMs found in the whole emitting region, although the
FWHMs slightly increase towards the southeast and decrease towards the west
from  the emission peak position. Such large line widths might be indicating
turbulent gas motions, which are more important close to the position proposed by
\citet{Dav01, Dav03} for an unseen embedded exciting source. Another
possibility is that the high FWHM values derived were due to the contribution
of several kinematic components of the line profiles. We shall analyse this
possibility later by using a multicomponent line profile model to fit the \sii\
profiles.

\begin{figure}
\centering
\includegraphics[angle=-90,width=0.95\hsize]{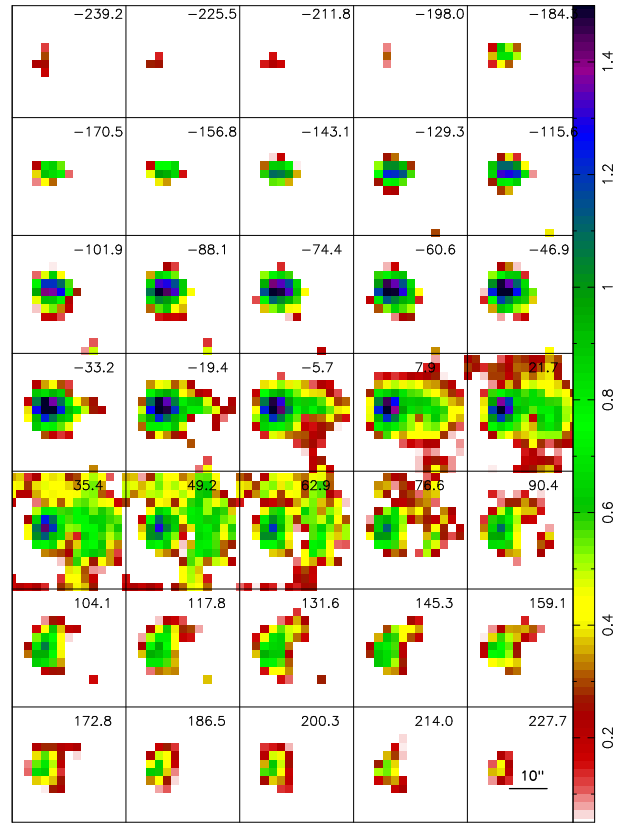}
\caption{
H$\alpha$ channel maps centred on the  \vlsr\  radial velocities 
(km~s$^{-1}$)  labelled in each panel. The spatial scale is given in one of the
panels. Fluxes  have been displayed on a logarithmic scale, in  units of
$10^{-16}$ erg~s$^{-1}$~cm$^{-2}$. (Channels corresponding to
$+21.7\le\mbox{\vlsr}\le+62.9$ are affected by residuals of an inaccurate
subtraction of the sky background at the north of the emission).
\label{fig6}}
\end{figure}

\subsubsection{Physical conditions} 

Line-ratio maps of the integrated line fluxes are suitable for getting spatial
information about excitation conditions (from \sii/H$\alpha$, \nii/H$\alpha$)
and electron density (from \sii\ 6717/6731) of the emitting gas. For the entire
region covered by the compact bright HHL~73 emission, the derived \sii\
6717/6731 line ratios go from $\sim0.6$ to $\sim0.9$, indicating a range of
electron densities $n_\rmn{e}=900$--4000 cm$^{-3}$, the highest values being
reached around the emission peak position. Electron density lowers towards the
west in the low-brightness collimated emission tail (\eg\ \sii\ 6717/6731 line
ratio $\sim1$, corresponding to $n_\rmn{e}\simeq600$ cm$^{-3}$, is derived at a
position $\sim5$ arcsec west of the peak position). These values were derived
for $T_\rmn{e}=10^4$~K.

The derived \sii/H$\alpha$, \nii/H$\alpha$ line ratio distributions show a
similar pattern. Both line ratios peak around the emission peak position, with
\sii/H$\alpha\simeq0.9$ and \nii/H$\alpha\simeq0.4$. Line ratios decrease
steeply (excitation increases) southeast of the peak, while the decrease is
slower to the west along the low-brightness emission tail.

\begin{figure}
\centering
\includegraphics[angle=-90,width=0.95\hsize]{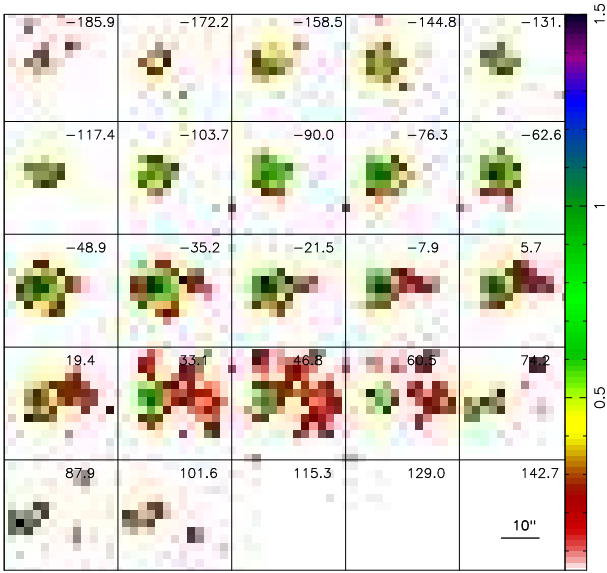}
\caption{
Same as Fig.\ \ref{fig6}, but for \nii\ $\lambda$6584~\AA.
\label{fig7}}
\end{figure}

\begin{figure}
\centering
\includegraphics[angle=-90,width=0.95\hsize]{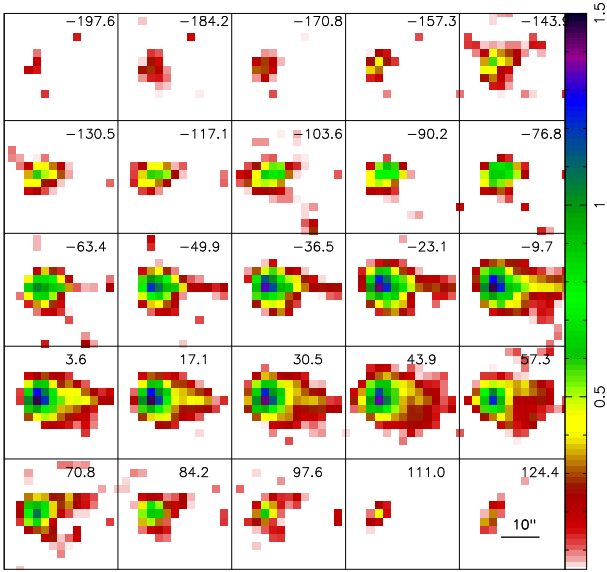}
\caption{
Same as Fig.\ \ref{fig6}, but for the \sii\ 6717~\AA\  emission line.
\label{fig8}}
\end{figure}

\subsection{IFS: Channel maps}

An inspection of the HHL~73 spectra at different positions revealed that the
line profiles, being broad and asymmetric,  display a variety of features
suggesting complex gas kinematics.

We will now analyse the behaviour of the emission as a function of velocity.
First, we obtained the H$\alpha$, \nii\ and \sii\ channel maps to explore the
3D structure of the physical conditions as a function of the radial velocity.
Furthermore, to analyse the complex line profiles found along the HHL~73
emission, we extracted spectra for each spaxel within the area covered by the
emission and we fitted the line profiles through a multicomponent line profile
model. 

\subsubsection{Spatial distribution of the emission}

\begin{figure}
\centering
\includegraphics[angle=270,width=0.8\hsize]{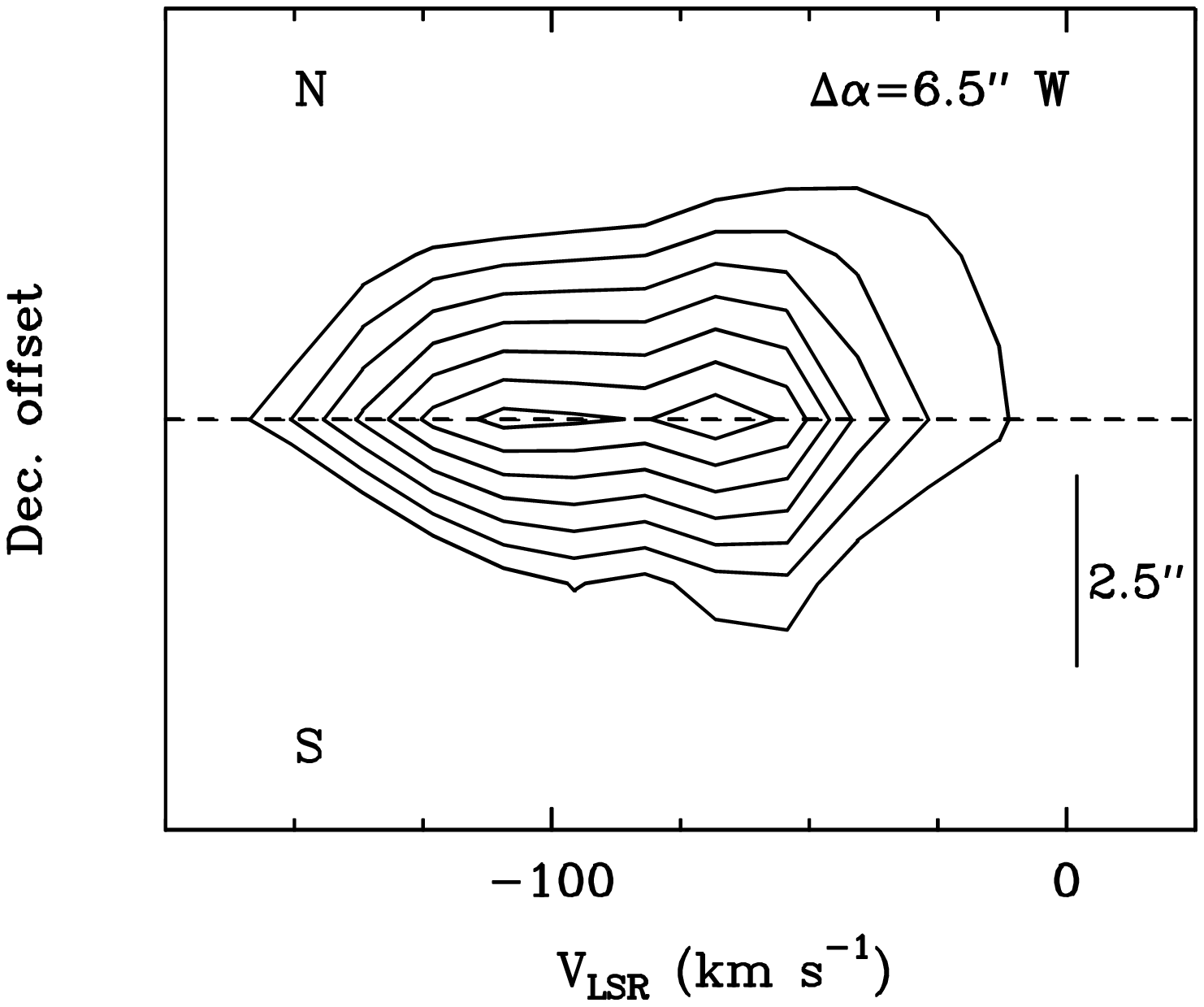}
\caption{
Position--Velocity diagram of the H$\alpha$ emission for a cut along the {\rm
Dec.} axis at $\rmn{RA}\simeq6.5$ arcsec west of the position of the intensity
peak. The $\Delta\delta=0$ position is marked by the dashed line. The contour
spacing is 10 per cent of the peak intensity. 
\label{fig9}}
\end{figure}

\begin{figure}
\centering
\includegraphics[angle=270,width=0.8\hsize]{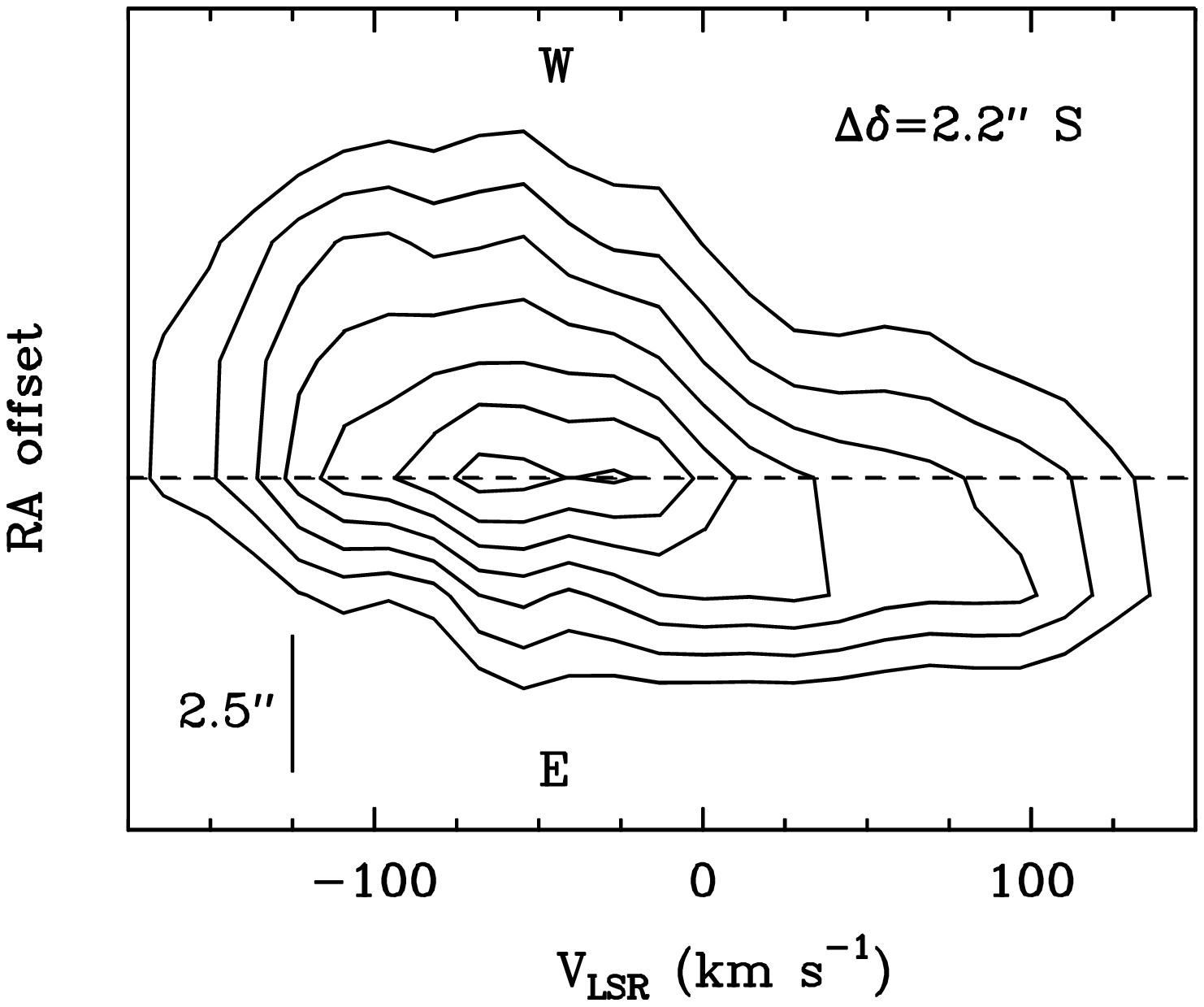}
\caption{
Position--Velocity diagram of the H$\alpha$ emission for a cut  along the {\rm
RA} axis at $\rmn{Dec.}\simeq2.2$ arcsec south of the position of the intensity
peak. The $\Delta\alpha=0$ position is marked by the dashed line. The contour
spacing is  10 per cent of the peak intensity.
\label{fig10}}
\end{figure}

A more detailed sampling of the gas kinematics was obtained by slicing the
datacube into a set of velocity channels along each emission line profile, thus
providing a ``tomograph'' of the emission line, not restricted to the centroid of
the line, but including also the wings. Each slice was made for a constant
wavelength bin of 0.3 \AA, corresponding to a velocity bin of $\sim14$
km~s$^{-1}$. The channel maps obtained for H$\alpha$, \nii\ $\lambda$6584 \AA\
and \sii\ $\lambda$6717 \AA\ are shown in Figs\ \ref{fig6}, \ref{fig7} and
\ref{fig8} respectively. As we used the same wavelength bin for all the lines,
the central velocity of a given channel map differs slightly from a line to
another. 

As can be seen from Figs\ \ref{fig6} and \ref{fig8}, the spatial distribution
of the H$\alpha$ and \sii\ emission found from the line-integrated IFS maps
(\ie\ the compact, bright region that extends towards the west in a
low-brightness and collimated tail) is also found in all the velocity channel
maps ranging from blueshifted \vlsr\ velocities of $-30$ km~s$^{-1}$ up to
redshifted velocities of $+20$~km~s$^{-1}$ in both emission lines.

The emission spreads over a wide velocity range for all the mapped lines, the
span of the velocity range being slightly different for each line.
The H$\alpha$ emission extends over the widest range as compared with the
other lines, covering a range 
$\sim 400$ km~s$^{-1}$ wide, from $-185$ to $+200$ km~s$^{-1}$. The strongest and
more extended H$\alpha$ emission is found for the velocity channels ranging from
$-50$ to $+20$ km~s$^{-1}$.
The \sii\ emission is detected within a velocity range  
$\sim 250$ km~s$^{-1}$ wide, from $-145$ to $+100$ km~s$^{-1}$. The strongest and
more extended \sii\ emission is found for the velocity channels ranging from
$-50$ to $+45$ km~s$^{-1}$. 
The \nii\ emission is detected within a velocity range 
$\sim 300$ km~s$^{-1}$ wide, from $-170$ to $+105$ km~s$^{-1}$. The \nii\ emission
presents a similar spatial extension in most of the velocity  channels since
the low-brightness emission tail is hardly detected in the \nii\ line.

Furthermore, we performed position--velocity ($PV$) diagrams at different 
right ascension and declination positions, both in the H$\alpha$ and \sii\
lines. Two of these H$\alpha$ cuts at selected positions are shown in order to
illustrate more clearly a trend found in the velocity distribution 
of the emission.

Figure \ref{fig9} shows the $PV$ diagram obtained for a cut at 
$\mbox{RA}\simeq6.5$ arcsec west of the position of the emission peak  (marked
in the figure by a dashed line). This cut along the declination axis has been
done at an RA position that intersects the beginning of the collimated emission
tail. The figure clearly shows that the overall emission appears  blueshifted
north and south of the emission peak, at $\mbox{\vlsr}\simeq-110$ km~s$^{-1}$.
Another velocity component is also found peaking at $\sim-70$ km~s$^{-1}$.
Furthermore, note that the spatial distribution of the velocities is slightly
asymmetric, with the range of velocities increasing from north to south: \eg\
the velocities range from  $-130$ to $-15$ km~s$^{-1}$ at $\sim2$ arcsec north
of the emission peak, while the velocities range from $-110$ to $-45$
km~s$^{-1}$ at $\sim2$ arcsec south of the emission peak.

Figure \ref{fig10} shows the $PV$ diagram obtained for a cut at 
$\mbox{Dec.}\simeq2.2$ arcsec south of the position of the emission peak. For
this cut along the RA axis, the figure shows that there is also a clear
asymmetry in the spatial distribution of velocities: the emission at blueshifted
velocities is spatially more extended towards the west, while at redshifted
velocities it is is more extended towards the east of the peak emission position. Two
blueshifted velocity components, peaking at $\sim-70$ and $\sim-25$ km~s$^{-1}$,
are also detected for this cut.

The inspection of different $PV$ diagrams obtained at several positions up to 
$\leq 10$ arcsec around the emission peak
revealed a general trend, both in the H$\alpha$ and \sii\ lines, in the highest
blueshifted velocities appearing towards the west of the emission peak position,
with the velocity values increasing from west to east along HHL~73.  
This signature of a two-velocity distribution in the
emission can be hardly appreciated in the channel maps.

\subsubsection{Density and excitation conditions}

We also explored  the behaviour of the density and excitation as a function of
the velocity by generating the line ratio channel maps for a  velocity range
from $-135$ to +110 km~s$^{-1}$.  

Regarding the kinematics of the emission, the \sii\ 6717/6731 line-ratio channel
maps indicate that $n_\rmn{e}$ decreases with increasing velocities: the
blueshifted gas has higher $n_\rmn{e}$ than the redshifted gas. This trend is
found at different position along the HHL~73 emission. Around the position of
the emission peak, we derived $n_\rmn{e}\simeq2200$ cm$^{-3}$ for the $-80$
km~s$^{-1}$ velocity channel, while $n_\rmn{e}\simeq740$ cm$^{-3}$ is derived
for the $-55$ km~s$^{-1}$ velocity channel.

Regarding the spatial distribution of the emission, the \sii\ 6717/6731 line
ratio channel maps indicate that $n_\rmn{e}$ decreases from east to west for all
the velocity channels, the $n_\rmn{e}$ variations being different depending on
the velocity. In fact, the variations found for $n_\rmn{e}$ along the east--west
axis seem to decrease with increasing velocity: \eg\ in the $-80$ km~s$^{-1}$
channel, $n_\rmn{e}$ changes from $\sim4200$ to $\sim740$ cm$^{-3}$ (\ie\ a
factor of $\sim5$) as we move $\sim10$ arcsec from east to west; in contrast,
$n_\rmn{e}$ only changes from $\sim1400$ to $\sim490$ cm$^{-3}$ for the same
spatial displacement in the +82 km~s$^{-1}$ channel.

The \sii/H$\alpha$ line-ratio channel maps show an increasing trend  
(\ie\ decreasing excitation) with increasing velocity, from blueshifted
velocities up to a velocity of $\sim+40$ km~s$^{-1}$. For higher redshifted
velocities, line-ratio values decrease (\ie\ excitation increases). Thus the
behaviour derived from the line ratios is that the excitation is higher at high
absolute velocity values (both blueshifted and redshifted) than at low
(absolute) velocity values.

Regarding the spatial distribution, the general trend found for each velocity
channel is a decrease in the line ratios (\ie\ an increase of excitation) from
east to west. 

\begin{figure}
\centering
\includegraphics[width=0.8\hsize,clip]{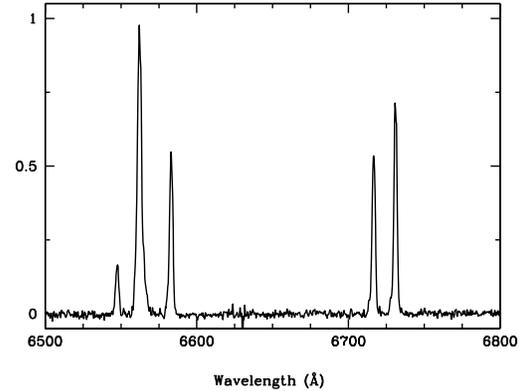}
\caption{
Spectrum extracted at the emission peak position. Flux has been normalized to
the flux of the H$\alpha$ peak. The characteristic HH emission lines
(H$\alpha$, \nii\ $\lambda$6548, 6583~\AA\ and \sii\ $\lambda$6717, 6731~\AA)
are clearly seen.
\label{fig11}}
\end{figure}

\subsection{IFS: Line-profile multicomponent analysis}

\begin{figure}
\centering
\includegraphics[angle=-90,width=\hsize]{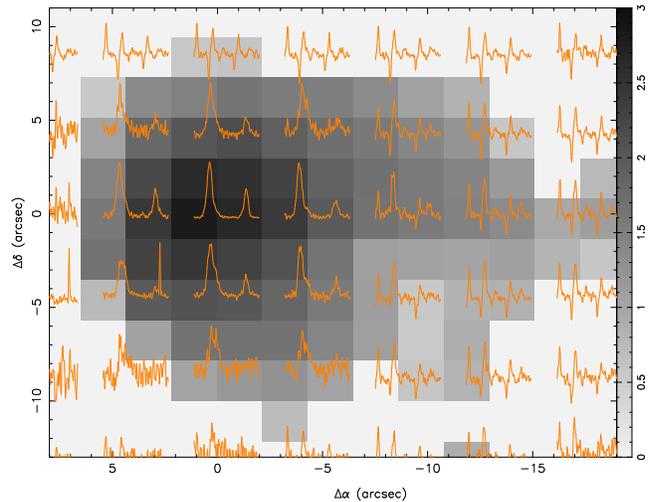}
\caption{
Two-dimensional distribution diagram of the observed spectra in the spectral
window 6553.1--6590.7 \AA, including the H$\alpha$ and \nii\ $\lambda$6584 \AA\
lines. Each spectrum has been obtained by averaging the spectra within an
aperture of $2\times2$ spaxels ($\sim4\times4$ arcsec$^2$) and its intensity has
been normalized to the H$\alpha$ peak. Spectra are overlaid on the H$\alpha$
integrated map obtained as in Fig.\ \ref{fig4}.
\label{fig12}}
\end{figure}

\begin{figure}
\centering
\includegraphics[angle=-90,width=\hsize]{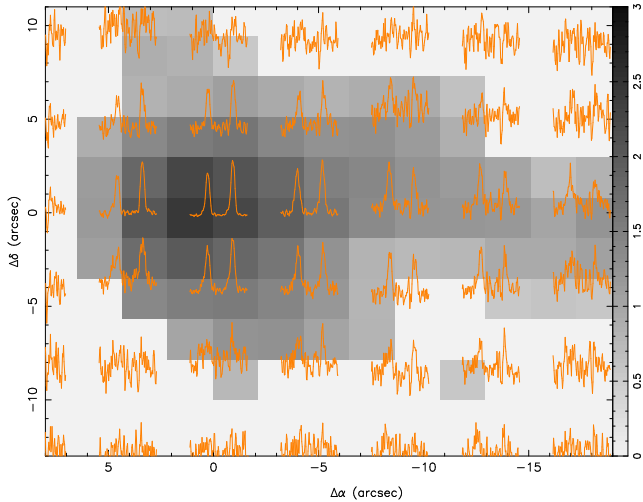}
\caption{
Same as Fig.\ \ref{fig13}, but for the \sii\ emission. The spectral window
selected for this plot is 6706.4--6739.5 \AA.
\label{fig13}}
\end{figure}

Figure \ref{fig11} shows the spectrum extracted at the H$\alpha$ emission peak
position. As  can be seen in the figure, the spectrum is closely reminiscent of
the characteristic shock-excited HH spectra.  Note in addition that the line
profiles are broad and asymmetric, suggesting that multiple kinematic components
are contributing to the line profile. Furthermore, the inspection of the spectra
extracted  at different positions over  HHL~73 reveals that the line profiles
display a  variety of features. Around the emission peak, line profiles  are
asymmetric and broad, with a faint blue double peak suggesting the presence of 
several gaseous systems. To the north and south of the emission peak, the
profiles  clearly  have contributions from at least two components, showing
double peaks and/or broad wings. Towards the west of the emission peak, in the
low-brightness collimated emission tail, the line profiles are more complex,
showing blue and red shoulders as well as broad wings. All the emission lines in
the observed spectral range show similar features to those described  here, as
can be seen in Figs~\ref{fig12} and \ref{fig13}, where  the spectra for the
spectral ranges which include the H$\alpha$ and \nii\ $\lambda$6584 \AA\ (Fig.\
\ref{fig12})  and the \sii\ $\lambda\lambda$6717, 6731 \AA\ emission lines
(Fig.\ \ref{fig13}) have been overlaid on the flux maps obtained by
integrating the signal within the same spectral range.

The complex line-profile pattern appears in both the H$\alpha$ and in the \sii\
lines. We used the \sii\ lines  to perform a line-profile decomposition to
reproduce the profiles of the HHL~73 spectra, since the  \sii\ emission should
be less  contaminated from the emission of a putative, undetected source and
thus, should better trace the kinematics of the shock-excited  outflow gas. The
\sii\ line profiles  were well reproduced by assuming the contribution from
three kinematic components. The line-profile decomposition was obtained from a
model including three Gaussian components. The Gaussian fitting was performed
using the {\small DIPSO} package of the  {\small STARLINK} astronomical 
software. The wavelength difference between the two \sii\ lines was fixed
according to their corresponding atomic parameters. Furthermore, the same FWHM
of the Gaussian  in each of the two \sii\ lines was assumed for each 
component of the profile.

\begin{figure}
\centering
\includegraphics[width=0.8\hsize]{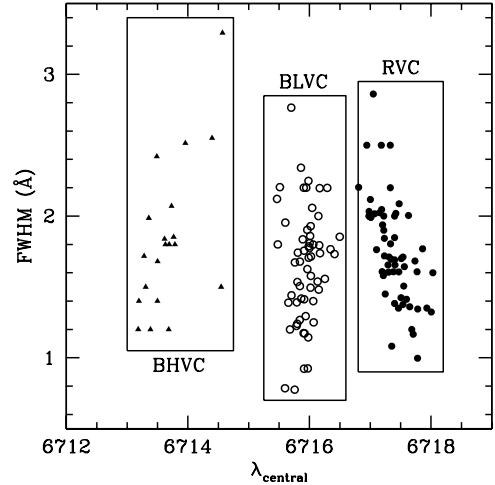}
\caption{
Central wavelength vs.\ FWHM for all the lines obtained by  fitting the HHL~73
spectra to a three Gaussian model. Symbols denote the so-called BLVC (open
dots), RVC (filled dots) and BHVC (filled triangles) velocity components (see
text).
\label{fig14}}
\end{figure}

\begin{figure}                         
\centering
\includegraphics[width=\hsize]{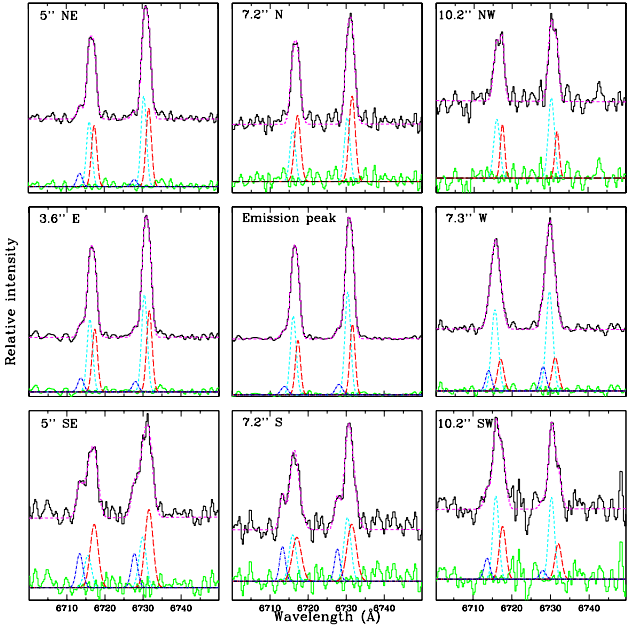}
\caption{
Observed spectra (black histograms) and fits (dashed magenta lines) obtained
from the model of three Gaussian components applied to the \sii\
$\lambda\lambda$6717, 6731 \AA\ emission lines. The individual components are
shown at the bottom of each panel. The same colour indicates the same component in
each line. Residuals obtained subtracting the fit from the observed spectra are
plotted in green. Labels in the left corner of each plot indicate the offset of
each observed spectrum from the \sii\ emission peak.
\label{fig15}}
\end{figure}

According to the fit parameters, we identified three Gaussians with three
different kinematic components well-separated in velocity (wavelength), as shown
in Fig.\ \ref{fig14}: two of the velocity components are blueshifted and the
third is redshifted. In the following, we refer to the blueshifted
high-velocity component as BHVC (filled triangles in the figure), to the
blueshifted low-velocity component as BLVC (open dots) and to the redshifted
velocity component as RVC (filled dots). Reasonably good results were obtained
from this three-Gaussian components model along the whole HHL~73 emission (see
Fig.\ \ref{fig15} for some examples, where the fits obtained at selected
positions and their residual are plotted). 
The \vlsr\ velocities of the BHVC range from $ -140$ to $ -90$ km~s$^{-1}$, with
values increasing towards the south. At the peak of the emission, the BHVC
reaches a value of $-100$ km~s$^{-1}$. 
The velocities of the BLVC range from $-50$ to $-2$ km~s$^{-1}$, with $-20$
km~s$^{-1}$ at the peak of the emission, with values increasing towards the
east.  
Finally, the velocities of the RVC range from $+20$ to $+40$ km~s$^{-1}$, with
$+35$ km~s$^{-1}$ at the peak of the emission. The highest values of the RVC are
found around the emission peak, and the velocity decreases towards the edges of
HHL~73.

\begin{figure}
\centering
\includegraphics[angle=-90,width=1.1\hsize]{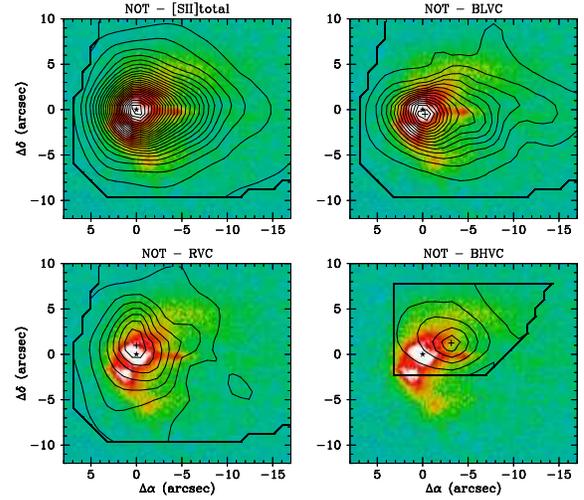}
\caption{ 
\emph{Top-left:} \sii\ intensity maps obtained from the fit of the 
three-Gaussian model to the emission line profiles and adding the individual fluxes
obtained for each velocity component. 
\emph{Top-right:} \sii\ intensity map for the BLVC. 
\emph{Bottom-left:} \sii\ intensity map for the RVC.
\emph{Bottom-right:} \sii\ intensity map for the BHVC. 
The intensity of each velocity component is overlaid on the narrow-band \sii\
NOT image.
\label{fig16}}
\end{figure}

The line-profile decomposition also shows significant differences, both in
the spatial distribution and in the relative strength of the emission
associated with each of the three velocity components (see Fig.\ \ref{fig16},
where the flux contours of each kinematic component are overlaid on the 
\sii-band NOT image). Regarding the spatial distribution, the BLVC and RVC
components are present in almost all the spectra, while the BHVC component is
concentrated in a smaller region, west of the emission peak. Regarding the
relative strength of the components, the emission from the BHVC is the weakest
(with flux values roughly one third of those found in the other components). 
The strongest emission comes from the BLVC component. Finally, the emission
with the widest spatial distribution comes from the RVC, the strength of its
emission being slightly weaker than that of the BLVC. 

Interestingly, we found a significant spatial displacement among the positions
of the emission peaks of the three velocity components (see also Figure
\ref{fig16}). The emission peaks of the BLVC and RVC components are closely
coincident (the estimated offset between the peaks of both components is
$\sim1.3$ arcsec; note that the IFS pixel size is 2.16 arcsec). In contrast, 
the BHVC peaks $\sim3.7$ arcsec northwest of
them. As can be seen in Fig.\ \ref{fig16}, the emission from the redshifted
component closely reproduces the spatial distribution of the bright bow-shaped
HHL~73 emission, while its contribution to the collimated, low-brightness
emission tail is lower than those from the blueshifted components. The two
blueshifted components appear mainly associated with the northwestern component
of the bright bow-shaped HHL~73 emission  and with the HHL~73 collimated
emission tail. The shape of the the BLVC
emission appears elongated in the east--west direction, extending towards the
west, and its emission peak closely coincides with the peak of the \sii\
emission found in the narrow-band NOT images. Finally, the emission from the
BHVC, located in a smaller area, also appears slightly elongated in the
east--west direction, peaking $\sim4$ arcsec west of the \sii\ emission peak. 
Note however that a clear association of each kinematic component with a given
morphological component of the HHL~73 \sii\ emission cannot be easily derived
only by looking to the \sii\ channel maps, since the emission from the wings of the
stronger BLVC and RVC emissions give an appreciable contribution to the emission 
at positions where the BHVC 
(the kinematics
component with the weakest emission) is located (\eg\ west of 
the peak of the emission), thus
contaminating the global emission at these positions.
Finally, it
should be pointed out that similar kinematic structures consisting of two
blueshifted (low and high) velocity components have already been detected in
Class~I YSO (\citealt{Tak06}) and in microjets from T Tauri stars
(\citealt{Bac00}; \citealt{Lav00}). 

\section{Discussion}

\subsection{The origin of the HHL~73 optical emission}

In all the positions covered by the HHL~73 emission, the spectra extracted from
the IFS datacube appear as low-ionization, line-emission spectra with no
appreciable continuum emission, (see Figs\ \ref{fig12} and \ref{fig13}). In
particular, the spectrum shown in Fig.\ \ref{fig11}, corresponding to the
emission of the $\sim2$~arcsec region around the peak position,  shows
no significant continuum emission superposed on the observed lines.

We compared the \nii/\sii\ and \nii/H$\alpha$ line ratio values measured at
several positions around the peak emission (where the signal-to-noise ratio is
high enough to do the diagnostic with confidence) with those obtained from the
diagnostic diagrams derived from the models of \citet{Kew02}. These models give
the \textit{ionization parameter} (\ie\ ionizing photon flux through an unit area)
of the photoionized gas found in \hii\ and star-forming regions as a function of
the line-flux ratios for a set of metallicities. The values found for the HHL~73
\nii/\sii\ and \nii/H$\alpha$ line ratios cannot be simultaneously fitted by the
photoionization models: the \nii/H$\alpha$ ratios are compatible with a low {\it
ionization parameter} for metallicities  $z\ge0.5\,z_\odot$; in contrast, for a
wide range of \textit{ionization parameters}, the \nii/\sii\ ratios are only
compatible with metallicity ranges significantly lower ($0.2\le
z\le0.5\,z_\odot$). Thus, it does not seem plausible to interpret the HHL~73
optical emission as mainly originating from photoionization from a stellar source. 

In contrast, the \nii/H$\alpha$ and \sii/H$\alpha$ line ratio values derived in
HHL~73 are fully compatible with those found for HH objects (\ie\ from
shock-excited emission) in the current diagnostic diagrams such as those of 
\citet{Can81}. Both HHL~73 line ratio values correspond to diagram regions 
where the HH objects are placed, being thus outside the regions where the spectra
from photoionized nebula, such as the planetary nebula and the \hii\ regions, are
located. It should be noted that the values measured for these two line ratios
(0.4--0.6 for \nii/H$\alpha$ and 0.9--1 for \sii/H$\alpha$) are more compatible
with the values found in HHs of high/intermediate excitation spectra than with
the values found in HHs of low-excitation spectra. The degree of excitation of
the HHL~73 spectra cannot be properly settled without data from the emission in
a bluer wavelength range including high excitation emission lines (\eg\
[O\,{\sc iii}]). Note, however, that the HH objects in which centimetre continuum
emission was reported currently show a high-excitation spectrum.

Thus, the HHL~73 IFS data suggest that the emission arises mainly from
shock-excited gas, characteristic of an HH object.

The contribution from continuum emission, if any, should be very weak, as derived from our
optical data. 
We  generated continuum images from the IFS datacube  by integrating, at each
position, the signal
over 
the  wavelength range free of lines emission. In these images,
we have only been able to find hints of
an emission enhancement relative to the neighbouring image background around the
HHL~73 position, but a at very low-level ($<$ 3$\sigma$).
In addition, the narrow-band, off-line filter NOT image fail
to detect continuum emission, although it has less depth than the IFS data.
This putative continuum contribution could
arise from light scattered by dust better than from a point-like stellar source,
as it extends over several pixels.

\subsection{The origin of the \nir\ emission associated with HHL~73}

Let us discuss the nature of the \tmas\ source found near the peak position of
HHL~73.

Regarding its size, the \tmas\ \textit{J}-, \textit{H}- and \textit{K}-band images show
that the \tmas\  source appears slightly extended when compared with other
sources in the fields. Typical FWHM of the field point sources are $3.3\pm0.2$ 
arcsec, while the FWHMs of the \tmas\ source are 5.7, 5.3 and 4.8 arcsec in the
\textit{J}-, \textit{H}- and \textit{K}-bands respectively, corresponding to a
deconvolved size of $\sim4$ arcsec ($\sim3500$ AU for a distance of 900 pc).

Regarding the \nir\ colour, the $(J-H, H-K)$ colour--colour diagrams are commonly
used to discriminate between pre-main sequence objects and field (main sequence
and giant) stars (see \eg\ \citealt{Ojh04}). The \nir\ colour indices of the
\tmas\ source derived from the \tmas\ photometry, $(H-K)=1.26\pm0.09$,
$(J-H)=1.47\pm0.11$, place the source inside the area corresponding to YSOs of
Class~II, whose \nir\ emission originates both from a photosphere and a
circumstellar disc, but its colours are also compatible, within the photometric
uncertainties, with being a Class~I source, having thermal emission from a
circumstellar envelope. Thus, from the \nir\ colours, it seems plausible to
infer that the \tmas\ source corresponds to a very red, embedded Class II or a
Class I source (\ie\ a typical outflow-driving source). 
According to current evolutionary models of PMS stars (\eg\ \citealt{Pa93};
\citealt{Sie00}),
the \tmas\ source lies in the region of the $(J,J-H)$ colour-magnitude diagram
that corresponds to low-mass YSOs with masses $\le1M_\odot$.
The size derived for
the \tmas\ source is compatible with that expected for an evacuated cavity
around a YSO. However, in this scenario the bulk of the \nir\ emission should mainly
arise from continuum scattered starlight, and this does not seem to be
the case. \citet{Dav01, Dav03} obtained high-resolution long-slit spectra at
the HHL~73 position and were unable to detect continuum emission in the {\it
H} and \textit{K} bands including \fe\ (at 1.64 $\mu$m) and H$_2$ (at 2.12 $\mu$m)
emissions. Thus, the continuum \nir\ emission in HHL~73, if present, is weak. 

An alternative origin for the bulk of the \nir\ emission is shock emission
instead of scattered starlight. In this case, the \tmas\ 
\textit{K}-band luminosity would mainly arise from the H$_2$ 2.12 $\mu$m line, the 
\textit{H}-band luminosity from the \fe\ 1.64 $\mu$m line, and the 
\textit{J}-band luminosity from other lines, like the \fe\ 1.25 $\mu$m line. 
However, this is hard to confirm, since no spectrum covering the \textit{J}-band
range is available, and the \tmas\ \textit{K} and \textit{H} magnitudes cannot be
compared reliably with the magnitudes derived from the line fluxes given by
\citet{Dav01, Dav03} -- the line fluxes correspond to the integrated emission
within a $0.8\times2.7$ arcsec, \ie\ only a small fraction of the source size.
In addition, the colour indices derived for the \tmas\ source are also
compatible with those expected for reddened molecular shocks, as was modelled by
\citet{Smi95} for a  wide range of shock parameters of molecular C- and
J-shocks, who found that  embebded shocks could mimic the \nir\ colours derived
for Class~0 sources, both sharing the same area of the \nir\ colour diagrams. 
Furthermore, \citet{Dav01, Dav03}, from the analysis of the kinematics derived
from the \nir\ spectra, suggest that the \fe\ and H$_2$ emissions are tracing
an outflow lying close to the plane of the sky, with a blueshifted component
towards the west and a redshifted component towards the east.

Thus, we found no conclusive evidence for the bulk of the \nir\ emission of
the \tmas\ source close to the HHL~73 peak to be mainly tracing the scattered
light from the YSO driving the outflow. Alternatively, a shock origin for the
bulk of the \nir\ emission seems to be compatible with the observations.

\subsection{The origin of the centimetre emission associated with HHL~73}

Centimetre continuum emission arising from HH objects has been detected in both 
high-mass (\eg\ HH 80-81: \citealt{Rod89}; \citealt{Mar93}, and
\iras~16547$-$4247: \citealt{Ga03}; \citealt{Rod05}) and low-mass star-forming
regions (\eg\ HH 1 and HH 2: \citealt{Pra85}; \citealt{Rod90}, and HH 32A:
\citealt{Ang92}). In high-mass star-forming regions, the centimetre emission
has a negative spectral index, which is consistent with non-thermal synchrotron
emission arising from shocks resulting from the interaction of a collimated
stellar wind with the surrounding medium, while in low-mass star-forming
regions the centimetre emission shows a flat spectrum, which is consistent with
optically thin free--free thermal emission from ionized gas. 

The spectral index of the centimetre HHL~73 emission cannot be derived from the
present single-frequency observations. Further observations at another
frequency are needed to estimate the spectral index in order to discriminate
between a thermal and non-thermal origin for the centimetre emission of HHL~73.
However, the exciting source of HHL~73 should be a low-/intermediate-mass YSO,
since its bolometric luminosity is low (see next section). Thus, the centimetre
emission detected in HHL~73 is most probably free-free thermal emission from
ionized gas.

Two mechanisms could ionize the gas: photoionization or shocks. The observed 
flux density (Table \ref{tjcmt}), at the distance of HHL~73, corresponds to a
centimetre continuum luminosity of 0.21 mJy~kpc$^2$. As discussed by 
\citet{Ang95}, the bolometric luminosity required to produce such a centimetre
continuum luminosity by photoionization is $\sim10^3~L_\odot$, more than an
order of magnitude higher than the upper limit derived for the HHL~73 exciting
source. In contrast, the HHL~73 centimetre continuum luminosity is typical of
thermal radio-jets tracing molecular-outflow exciting sources, whose
radio-continuum emission is produced by shock ionization. In addition, the
momentum rate of the HHL~73 molecular outflow, derived from the CO data of
\citet{Dob93}, fits the momentum rate vs.\ centimetre continuum luminosity
relationship expected for shock-ionization induced emission in
molecular-outflow sources of low bolometric luminosity \citep{Ang95}.

In conclusion, the centimetre continuum emission of HHL~73 most probably
arises from shock-ionized gas.

\subsection{Is \iras~21432+4719 tracing the exciting source of HHL~73?}

\iras\ sources in star-forming regions are frequently found tracing the driving
source of molecular and atomic outflows. The \iras\ source closest to HHL~73 is
21432+4719, whose position from the \iras\ Point Source Catalogue is  $\sim30$
arcsec northeast of the HHL~73 peak position ($\sim0.13$ pc for a distance of
900 pc), with a large uncertainty, so that HHL~73 lies outside the \iras\ error
ellipse, near its southwestern border (see Fig.\ \ref{fig1}).

\iras~21432+4719 is well detected in three \iras\ bands, having a rising
far-\textit{IR} spectral energy distribution, with flux densities of 1.42,
12.19 and 23.45 Jy at 25, 60 and 100 $\mu$m respectively, and an upper limit of
0.25 Jy at 12 $\mu$m. The bolometric luminosity derived from the \iras\ flux
densities, following the recipe of \citet{Con07}, is 27 $L_\odot$.  This
luminosity corresponds to a ZAMS star of 2.6 $M_\odot$. The \iras\ colours
derived from the fluxes are typical of embedded protostellar sources
(\citealt{Bei86}; \citealt{Dob92}), being redder than those found for T
Tau-like sources (\citealt{Har88}). 

Let us discuss whether \iras~21432+4719 is tracing the embedded exciting source
of HHL~73.

As can be seen in Fig.\ \ref{fig1}, HHL~73 lies close to the border of a region
of high visual extinction. A detailed study of the brightness distribution of
the optical and \nir\ emission of HHL~73 shows that there is a shift towards
the northeast of the emission centroid position with increasing wavelength. 
The shift between the H$\alpha$ and \textit{K}-band emission centroids is
$\sim1.5$ arcsec, while between the \textit{J}- and \textit{H}-, and between
the {\it H}- and \textit{K}-band emission centroids it is $\le1$ arcsec. These
shifts are indicative of a large gradient of extinction, with extinction
increasing towards the northeast. Thus, if there is \fir\ emission associated
with HHL~73, we expect to find it offset to the northeast of the optical and
\nir\ emissions.

As already mentioned, the catalogue position of \iras~21432+4719 is offset from
HHL~73 in the expected direction, although somewhat displaced from the HHL~73
position. Nonetheless, the region encompassed by the \iras\ error ellipse, \ie\
the region with the highest probability of finding the \iras\ source, almost
reaches the HHL~73 location. Thus, the \fir\ emission could be located nearer
to the optical and \nir\ emission than the \iras\ catalogue position. 

No source is detected in the area encompassed by the \iras\ error ellipse in
the four bands of the \textit{Midcourse Space Experiment (MSX)}, at wavelengths
from 8 to 22 $\mu$m. However, it should be noted that at 8.28 $\mu$m, in the
highest-sensitivity \textit{MSX} band, there are hints of an enhancement in
emission over the local background in a few pixels covering the position of the
\tmas\ source, but at a level below 3 times the image rms.

In conclusion, in spite of the angular separation between the \iras\ source and
the optical and \nir\ emission of HHL~73, \iras~21432+4719  is the most plausible
candidate to  be driving  HHL~73, although the evidence is inconclusive and new
observations in the millimetre wavelength range are needed to confirm this
issue.

\subsection{A speculative scenario for HHL~73}

\begin{figure}                         
\centering
\includegraphics[width=0.8\hsize]{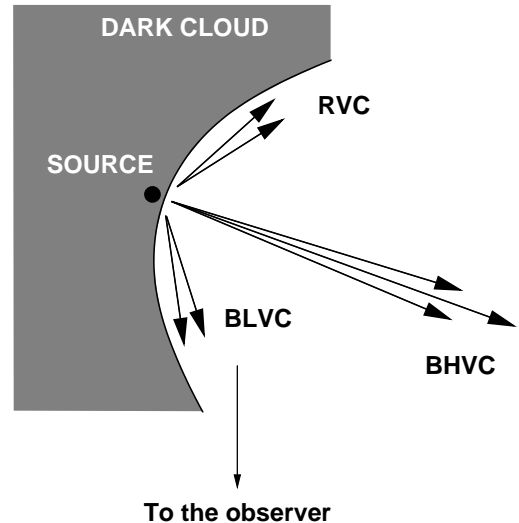}
\caption{Sketch of the scenario proposed for HHL 73.
\label{fig17}}
\end{figure}

The HHL~73 morphology is reminiscent of the HH~83 system (\eg\
\citealt{Rei89}). \iras~05311$-$0631, the proposed exciting source of HH 83, is
found located within a region with high visual extinction, close to the apex of
the optical emission. The HH~83 system is interpreted as an optical jet
emerging from a conical cavity evacuated by the outflow. The counterjet is not
visible since it flows inside the high-extinction region. The bulk of the 
bow-shaped
emission at the base of the HH~83 jet arises from reflected light of the
driving source, tracing the walls of the cavity. Apparently, the HH~83 system
could mimic the HHL~73 scenario. 

However there are significant differences between both objects. 
First, the bulk of the HHL~73 emission detected is line emission, with small,
if any, continuum contribution. In contrast, the emission from the bow-shaped
base of the HH~83 jet  has an appreciable continuum contribution. 
Second, the radial velocities measured in HHL~73 present both redshifted and
blueshifted values, while the radial velocities along the HH~83 jet are only
blueshifted. Thus, a different scenario with another geometry than that
proposed for the HH~83 system should be assumed to explain the spatial
distribution of the HHL~73 velocities. 
In addition, the separation between \iras~21432+4719 and the optical and \nir\ 
emission of HHL~73 is appreciably larger than that between \iras~05311$-$0631 
and the base of the HH~83 jet ($\sim13$ arcsec or $\sim0.03$ pc for a distance
of 460 pc). However, as already discussed, \iras~21432+4719 could be located
closer to the apex of HHL~73 than the  position given by the \iras\ Point
Source Catalogue.

Let us describe a qualitative scenario compatible with the complex
HHL~73 kinematics derived from IFS data. Let us assume that the YSO exciting
HHL~73 is deeply embedded and located to the northeast of HHL~73,
near the apex of the HHL~73 emission.  
This source is ejecting supersonic gas with a low-collimation pattern (as,
\eg, from a disc- or X-wind) and, in addition, is driving a highly-collimated
outflow (see, for example, \citet{Tak06} and references therein, where several
models
giving rise to outflows having low- and high- velocity components are
discussed). 
The low-collimated gas flowing outside the dark cloud can give, by projection
effects, both the low-blueshifted and the redshifted observed velocities (\ie\
the main contributors to the BLVC and RVC), while the BHVC would mainly arise
from the high-collimated outflow. 
The global kinematic pattern obtained from such a scenario would then mimic
that produced by an outflow having a redshifted and two (low and high)
blueshifted velocity components, even if all the supersonic gas we are
observing is actually flowing outside the dark cloud (see the sketch in Fig.\
\ref{fig17}).

\section{Conclusions}

From the analysis carried out of the CCD and IFS HHL~73 observations,
complemented with those available from the literature we found:
\begin{description}

\item 
The bulk of the HHL~73 optical red emission is mainly line emission from
shock-excited gas, such as is usually found in HH objects.  Continuum
emission from dust scattered starlight, if present, seems to be weak.

\item 
High-spatial resolution, narrow-band images of HHL~73 resolved its 
bow-shaped morphology into two components and a collimated emission $\sim4$ 
arcsec long emerging towards the west, which is reminiscent of a microjet.

\item 
The kinematics of HHL~73 derived from the IFS data show that the emission is
supersonic, with both blueshifted and redshifted velocities. From a line-profile
multicomponent model, we were able to resolve three velocity components for the
\sii\ $\lambda\lambda$6717, 6731 \AA\ emission lines. These kinematic components
correlate with the morphological components discovered from the high-resolution
CCD images. The redshifted component, being the most extended one, spatially
correlates with the bright, extended HHL~73 emission, elongated  in the north--south
direction. Its spatial distribution is also in good agreement with that of the
radio continuum emission. The two blueshifted (low- and high-velocity)
components are morphologically well correlated with the northwestern HHL~73
emission and the microjet. In order to mimic the complex kinematic pattern
derived from the IFS spectra, we tentatively propose a scenario involving both
low- and highly-collimated supersonic gas ejected from a YSO.

\item
In the light of the data here discussed, we propose that the exciting source of
HHL~73 (and possibly of HH 379) is deeply embedded in the dark cloud causing
the high extinction, lying to the northeast of HHL~73, close to its apex.
In spite of its angular separation, \iras~21432+4719 is the most plausible
candidate to be tracing the
low-/intermediate YSO that is driving the centimetre, \textit{IR} and optical
emission associated with HHL~73, although high sensitivity observations with
high angular resolution in the millimetre wavelength range, not affected by
extinction, are needed to confirm this issue and to establish clearly the
location and properties of the HHL~73 driving source.

\item
Further observations, both in the centimetre
wavelength range and IFS data covering the wavelength range blueward of
H$\alpha$, to include high-excitation HH emission lines, are needed to
settle the nature of the HHL~73 object completely.
\end{description}

\section*{Acknowledgments} 

We acknowledge the Support Astronomer Team of the IAC for obtaining the NOT
images.
We thank Aina Palau for help with the MSX data.
We appreciate Terry Mahoney's help with the manuscript.
The authors were supported by the Spanish MEC grants 
AYA2005-08523-C03-01 (RE, RL, AR and GB),
AYA2005-09413-C02-02 (SFS),
AYA2006-13682 (BGL) and 
AYA2005-08013-C03-01 (AR).
In addition we acknowledge 
FEDER funds (RE, RL, AR and GB), and
the Plan Andaluz de Investigaci\'on of Junta de Andaluc\'{\i}a as research group
FQM322 (SFS). 
ALFOSC is owned by the Instituto de Astrof\'{\i}sica de Andaluc\'{\i}a (IAA) and
operated at the NOT under agreement between IAA and the NBIfAFG of the
Astronomical Observatory of Copenhagen.
This publication makes use of data products from the Two Micron All Sky Survey,
which is a joint project of the University of Massachusetts and the Infrared
Processing and Analysis Center/California Institute of Technology, funded by the
National Aeronautics and Space Administration and the National Science
Foundation, and data products from the Midcourse Space Experiment (MSX).
We thanks the anonymous referee for his/her useful information and comments.

\bsp

\label{lastpage}

\end{document}